\documentclass[aip,jcp,11pt,citeautoscript,amsmath,amssymb,reprint]{revtex4-1}

\newif\ifpdf\ifx\pdfoutput\undefined\pdffalse\else\pdfoutput=1\pdftrue\fi

\usepackage{graphicx}
\usepackage[usenames]{color} 
\usepackage{comment} 
\usepackage{bm}
\newcounter{abc}

\newcommand{\be}{\begin{equation}} 
\newcommand{\ee}{\end{equation}}
\newcommand{\bea}{\begin{eqnarray}} 
\newcommand{\eea}{\end{eqnarray}}

\newcommand{\rv}{\bm{r}} \newcommand{\f}{f} 
 
\newcommand{\p}{^{(0)}} 

\newcommand{\effe}{{\mathcal{E}}}

\newcommand{\Rsetgamma}{\{ \vec{R}\}^{\gamma}}
\newcommand{\Rsetgammaprime}{\{ \vec{R}\}^{\gamma^{\prime}}}

\renewcommand{\vec}[1]{\bm{#1}}

\begin{document}

\title{\bf Phase behaviour of polydisperse spheres: simulation strategies
and an application to the freezing transition}

\author{Nigel B. Wilding} \affiliation{Department of Physics, University of Bath, Bath BA2
7AY, United Kingdom.}
\author{Peter Sollich} \affiliation{King's College London, Department of
Mathematics, Strand, London WC2R 2LS, United Kingdom.} 

\begin{abstract} 

The statistical mechanics of phase transitions in dense systems of
polydisperse particles presents distinctive challenges to computer
simulation and analytical theory alike. The core difficulty, namely
dealing correctly with particle size fractionation between coexisting
phases, is set out in the context of a critique of previous simulation
work on such systems. Specialized Monte Carlo simulation techniques and
moment free energy method calculations, capable of treating
fractionation exactly, are then described and deployed to study the
fluid-solid transition of an assembly of repulsive spherical particles described
by a top-hat ``parent'' distribution of particle sizes. The cloud curve
delineating the solid-fluid coexistence region is mapped as a function
of the degree of polydispersity $\delta$, and the properties of the
incipient ``shadow'' phases are presented. The coexistence region is
found to shift to higher densities as $\delta$ increases, but does not
exhibit the sharp narrowing predicted by many theories and some
simulations.

\end{abstract}

\maketitle 
\section{Introduction} 
\label{sec:intro}

A complex fluid is described as ``polydisperse'' when its constituent
particles are not all identical, but exhibit a spread of size, shape or
charge. Polydispersity is a pervasive feature of both natural and
synthetic macromolecular systems such as colloids, polymers and liquid
crystals. But for many years it was regarded as a practical and
conceptual nuisance to be minimised in experiment and ignored by theory.
More recently, however, it has become clear that polydispersity is
actually a matter of fundamental interest and practical importance in
its own right, giving rise as it does to a rich variety of phenomena not
observed in monodisperse systems (for a review, see Sollich\cite{Sollich2002}).
Examples include novel phase behaviour such as an extreme sensitivity of
phase boundaries to the presence of rare large particles
\cite{wilding2005b}, critical points that occur below the maximum
coexistence temperature\cite{kita1997a,Wilding08A},  density dependent
wetting transitions  \cite{Buzzacchi2006a} and surface size segregation
effects \cite{Pagonabarraga2000,Buzzacchi2004}, as well as interesting
dynamical effects such as an enhanced propensity to glass formation
\cite{Zaccarelli2009} and the possibility of multistage relaxation
processes \cite{Warren1999}. 

Despite the upsurge of interest in polydispersity induced phenomena in
complex fluids, several basic questions remain.  A prime example is the
nature of the freezing transition for simple spherical particles with a
spread of diameters, which in quantitative terms
is conveniently characterized by a parameter $\delta$
measuring the standard deviation of the diameter distribution in units
of its mean. Intuitively one expects that polydispersity should alter
the location of the freezing curve and destabilize crystalline phases
\cite{Pusey1987}. However, the character and extent of these alterations
remain unclear despite extensive study  by experiment
\cite{Pusey1986,Pusey1991}, density functional theories
\cite{Barrat1986,mcrae1988}, simplified analytical theories
\cite{Phan1998,Pusey1987,Bartlett1997,Sear1998, Bartlett1999,Xu2003} and
simulation \cite{Dickinson1981,Dickinson1985,Stapleton1990,Bolhuis1996b,
Kofke1999,Lacks1999,Huang2004,Fernandez2007,
Yiannourakou2009,Yang2009,Fernandez2010,Nogawa2010,Nogawa2010a}.  On the
experimental side there is evidence for a ''terminal'' degree of
polydispersity $\delta_t$ above which a fluid will not crystallize. This
is supported by a number of theoretical and simulation studies, some of
which suggest that the terminal polydispersity arises from a progressive
narrowing of the fluid-solid coexistence region with increasing
$\delta$, with the boundaries of this region meeting at a point of equal concentration
\cite{Phan1998,Bartlett1999,Huang2004,Nogawa2010,Nogawa2010a}. A reentrant melting
scenario has also been proposed whereby compressing a crystal with a
polydispersity slightly
below $\delta_t$ can cause it to melt\cite{Bartlett1999}. One simulation
study predicts the occurrence of a partly crystalline ``inhomogeneous
phase'' \cite{Fernandez2007} at high polydispersity, while other work
has argued that the suppression of crystallization is a dynamical effect
arising variously from the low diffusivity of large particles
\cite{Evans2001a}, the intervention of a glass transition
\cite{Pusey1987a,Chaudhuri2005,Fernandez2010}, or anomalously large
nucleation barriers \cite{Auer2001a}.

In our view, the reason for the diverse predictions of theory and
simulation regarding the equilibrium phase behaviour is that much
previous theoretical and simulation work has failed to cater properly
for a key feature of polydisperse phase equilibria, namely {\em
fractionation} -- the phenomenon whereby the distribution of particle
sizes can be different among coexisting phases, whilst the overall
distribution (across all phases) is fixed. One approach that does
incorporate fractionation exactly (within the context of a mean field
theory) is the moment free energy (MFE) method. Previous MFE
calculations by one of us \cite{FasSol03,Fasolo2004} find neither
evidence for narrowing of the coexistence region with increasing
$\delta$, nor reentrant melting. Instead they predict that a
polydisperse fluid can always split off a small amount of a solid phase
having a narrow distribution of particle sizes. The purpose of the
present paper is to compare the findings of MFE calculations with the
results of tailored Monte Carlo (MC) simulations that similarly cater
exactly for fractionation effects. 

The paper is arranged as follows. In Sec.~\ref{sec:phenomenology} we
outline principal aspects of the phenomenology of polydisperse phase
equilibria. We then present in Sec.~\ref{sec:simstrat} a detailed
discussion of general issues surrounding the best choice of simulation
ensemble for obtaining accurate estimates of phase coexistence
properties.  Sec.~\ref{sec:methods} describes the model system
we have studied and the bespoke techniques we have developed to
determine fluid-solid coexistence properties. The results of applying
these techniques to polydisperse spheres are described in
Sec.~\ref{sec:results}. Sec.~\ref{sec:discuss} summarizes our conclusions.

\section{Phenomenology of polydisperse phase equilibria}

\label{sec:phenomenology}

Typically one describes the polydispersity of a given system in terms of a
density distribution $\rho\p(\sigma)$ which counts the number of particles
per unit volume whose value of the polydisperse attribute $\sigma$, lies in
the range $\sigma\dots\sigma+d\sigma$ \cite{Salacuse1982}. In most real polydisperse systems
the form of $\rho\p(\sigma)$ is {\em fixed} by the synthesis of the fluid,
and only its scale can change according to the degree of dilution of the
system. Accordingly, one writes
\be
\rho\p(\sigma)=n\p f(\sigma)
\label{eq:dilution}
\ee
where
$f(\sigma)$ is a normalized fixed shape function and $n\p$ is the overall
number density. Varying $n\p$ corresponds to scanning a ``dilution line''
of the system.

At coexistence, particles with different $\sigma$ values will be
partitioned unequally between the phases. This is the phenomenon of
``fractionation''. To describe it, it is necessary to define separate
``daughter'' density distributions $\rho^{(\gamma)}(\sigma)$ ($\gamma=1,2,
\ldots$) which 
measure the distribution of the polydisperse attribute for each phase $\gamma$.
When the polydispersity is fixed, particle conservation implies that the
weighted average of the daughter distributions equals the fixed overall
density distribution, or ``parent'' $\rho\p(\sigma)$, i.e.
\be
\rho\p(\sigma)=n\p f(\sigma)=\sum_\gamma\xi^{(\gamma)}\rho^{(\gamma)}(\sigma),
\label{eq:lever}
\ee 
with $\xi^{(\gamma)}$ the fractional volume of phase $\gamma$, where $\sum_\gamma\xi^{(\gamma)}=1$.
This expression represents a generalisation of the lever rule to
polydisperse systems.

To illustrate the profound differences between phase behaviour in
monodisperse and polydisperse systems, it is instructive to recall first
the familiar case of the binodal curve of a monodisperse system in the
density-temperature plane. This curve serves a dual purpose: on the one
hand it describes the range of overall densities for which phase
coexistence occurs; and on the other hand it identifies the densities of
the coexisting phases themselves. Now, for a polydisperse system the range
of overall (parent) densities that leads to coexistence is similarly
delineated by a curve in $n\p$-temperature space -- the so-called ``cloud''
curve. However, the densities of the coexisting phases themselves do not in
general coincide with the cloud curve. Instead, as one varies the parent
density $n\p$ through the coexistence region at a fixed temperature (say),
one generates an infinite sequence of pairs of differently fractionated
coexisting phases.

\begin{figure}[h]
\includegraphics[type=pdf,ext=.pdf,read=.pdf,width=0.7\columnwidth,clip=true]{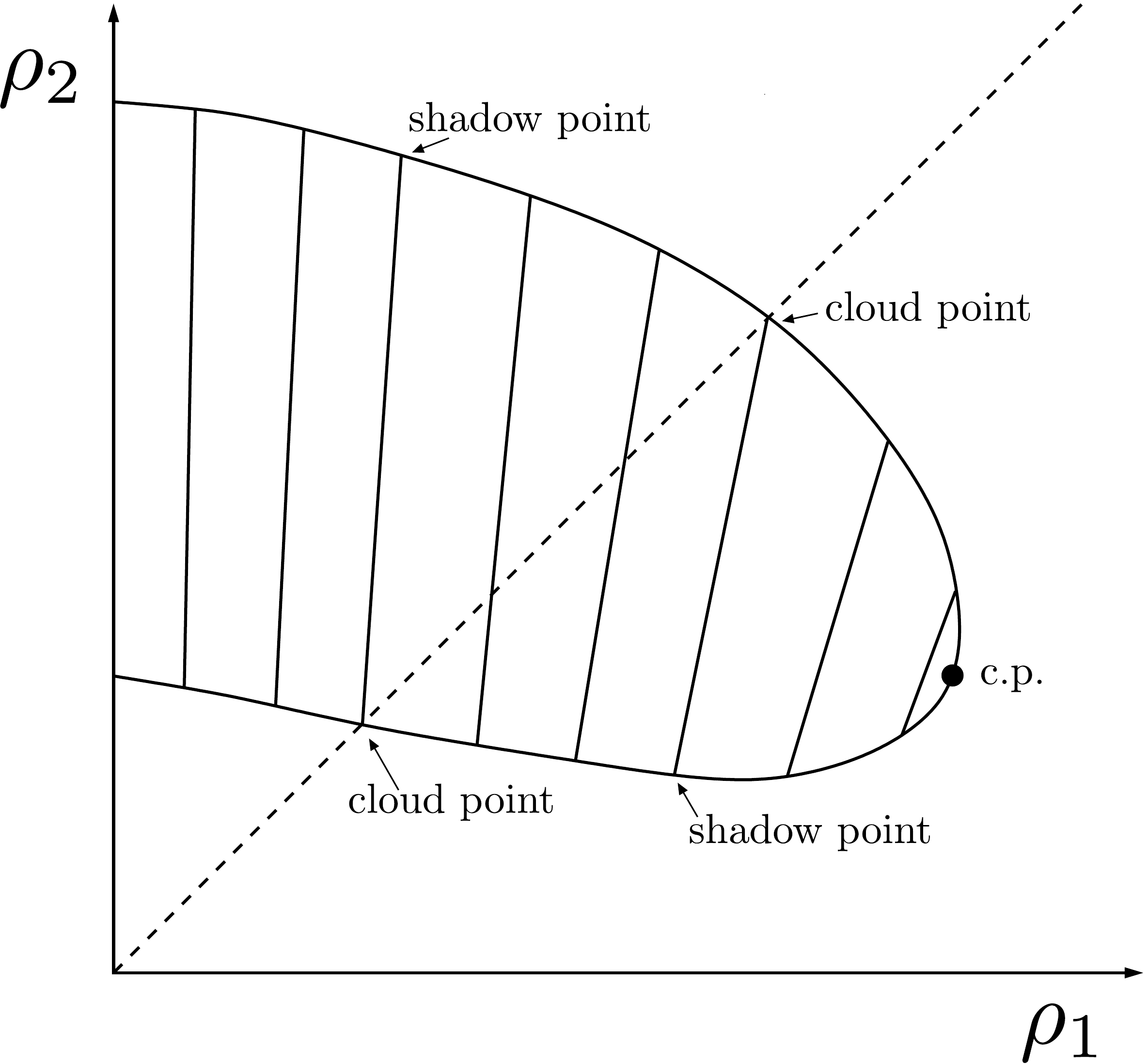}
\caption{Schematic of an isothermal cut through the fluid-fluid phase
diagram of the binary mixture described in the text. Thick solid line:
coexistence boundary; thin solid lines: tie-lines; dashed line:
dilution line, here chosen as $\rho\p_1=\rho\p_2$.} 
\label{fig:binary} 
\end{figure}

Insight into this phenomenology can be gained by considering the
simplest case of a bidisperse (binary) mixture with densities $\rho_1$
and $\rho_2$ of two species of particles of different sizes; these are
the analog of $\rho\p(\sigma)$. Fig.~\ref{fig:binary} sketches an
isothermal cut through an exemplary bulk phase diagram, showing a region
of fluid-fluid phase separation with tie-lines that shrink to zero at a
critical point (c.p.).\footnote{We assume that species 2 (taken as the
larger particles) has stronger attractive interactions and so phase
separates on its own (vertical axis) at the chosen $T$ while the pure
species 1 fluid (horizontal axis) does not. With interaction strengths
chosen in this way, species 2 particles will typically accumulate in the
denser phase, with its shorter interparticle distances. The tie-lines in
the sketch therefore cross any dilution line from below, moving from
smaller values of $\rho_2/\rho_1$ in the low density phase to larger
ones for the dense phase.} The dilution line constraint of a fixed shape
for $\rho\p(\sigma)$ reduces in the bidisperse case to a fixed ratio
$\rho\p_1/\rho\p_2$ (indicated by the dashed line in
Fig.~\ref{fig:binary}). As $n\p=\rho\p_1+\rho\p_2$ is increased from
zero, the system follows the dilution line into the coexistence region,
which it enters (and leaves) at a ``cloud point''. For a given point on
the dilution line inside the coexistence region, the parent splits into
two daughter phases located at the ends of the tie-line which intersects
this point. However, owing to fractionation, the daughters lie {\em off}
the dilution line.

Since an infinite number of points occupy the dilution line between the
cloud points, an infinite sequence of differently fractionated coexisting
states is encountered as the coexistence region is traversed. Customarily
one singles out the end points of this sequence for special attention, i.e.\
the case of incipient phase separation that occurs when the value of $n\p$
coincides with one of the two cloud points. Under these conditions, one
daughter phase has a fractional volume of essentially unity and
consequently (from the lever rule, Eq.~(\ref{eq:lever})) a density
distribution that is identical to the parent, while the other phase --
known as the ``shadow'' -- has an infinitesimal fractional volume and a
density distribution that deviates maximally from that of the parent. The
curve formed by plotting the number density (or packing fraction) of the
shadow phase as a function of temperature is known as the shadow curve.

Qualitative differences between the phase behaviour of monodisperse and
polydisperse systems are also evident in other projections of the full
phase diagram. For instance, in the pressure-temperature plane, coexistence
for a monodisperse system occurs along a line. By contrast for a
polydisperse system, coexistence occurs within a {\em region} of the
pressure-temperature phase diagram \cite{Rascon2003,bellier-Castella2000}.
This is because each of the infinite sequence of coexisting states along
the dilution lines is associated with a different pressure, and so the
pressure varies as one crosses the coexistence region. For a fuller
account of the phenomenology of phase behaviour in polydisperse fluids, we
refer the reader to a previous review \cite{Sollich2002}.

\section{Simulation strategies for polydisperse systems}
\label{sec:simstrat}

Computer simulation is a principal route to determining the phase
behaviour of model thermal systems, and many such investigations of
polydisperse systems have been reported.
\cite{Dickinson1981,Dickinson1985,Stapleton1990,Bolhuis1996b,
Kofke1999,Lacks1999,Huang2004,Fernandez2007,
Yiannourakou2009,Yang2009,Fernandez2010,Nogawa2010,Nogawa2010a} However
the results that have emerged from the various studies are generally
much more at variance with one another (even for a given model)
than is typically the case in comparable studies of monodisperse
systems. It therefore seems appropriate to consider carefully what
particular issues and challenges polydispersity presents for simulation
and how best to tackle them in practice. To
this end we provide here a detailed critical assessment of the relative
utility of the various statistical ensembles and simulation strategies
that have been employed previously for the computational study of phase
transitions in polydisperse systems.  Our discussion distinguishes two
categories of ensemble: those that fully permit density fluctuations at
the level of individual particle species (``unconstrained density
ensembles''), and those (``constrained density ensembles'') that do not.
We argue that the latter category are seriously deficient both with
regard to their ability to probe equilibrium phase behaviour, and their
susceptibility to finite size effects. By contrast, unconstrained
density ensembles -- whilst requiring some effort to fully realize their
benefits -- constitute the method of choice for accurate studies of
polydisperse phase equilibria.

\subsection{Unconstrained densities}


As is now well established \cite{Wilding1995,Panagio2000}, an efficient
and accurate strategy for determining phase boundaries in monodisperse
systems is to employ an ensemble that captures on a {\em global} scale
the fluctuations characteristic of the transition. Prime examples are
the grand canonical ensemble (constant $\mu,V,T$) and the
isobaric-isothermal ensemble (constant $N,p,T$) \cite{Frenkelsmit2002}.
The common feature of these ensembles is that they avoid physical
contact (i.e.\ interfaces) between coexisting phases even when operating
at a coexistence state point. Instead the two phases are linked -- and
their relative free energies measured -- via a phase space path
\cite{Bruce2003} (the negotiation of which may entail biased sampling)
that permits the simulation to fluctuate back and forth between the
disjoint configuration spaces of the pure phase states. Although this
path may traverse mixed phase (i.e.\ interfacial) states en-route between
the pure phases, the associated surface tension penalty is such that at
any one time, the system will be found with overwhelming probability in
{\em pure} phase states. Accordingly there is ample opportunity to
sample the properties of the phases whilst they occupy the maximum
available system volume -- thus minimizing finite-size effects. Indeed
it has been shown that when the system is found with equal probability
in each phase (the ``equal weight criterion''), finite-size corrections
to measured coexistence properties are exponentially small in the system
size \cite{Borgs1992}.

In seeking to study phase behaviour in polydisperse systems, it is
clearly desirable to retain the aforementioned benefits of the ($\mu,
V,T$) and ($N,p,T$) ensembles, whilst at the same time generalizing them
to permit sampling of the fluctuations in the polydisperse attribute.
More specifically, one wishes to cater for fractionation so that the
distribution of $\sigma$ (which for definiteness we shall take as the particle size) can
vary from phase to phase. For the particular case of MC simulations
within the ($\mu,V,T$) ensemble, the total particle number fluctuates
(by means of insertions and deletions) and incorporating polydispersity
entails imposing a distribution of chemical potentials $\mu(\sigma)$ and
additionally introducing particle resizing updates. Doing so permits
fluctuations not only in the overall instantaneous size distribution
$\rho(\sigma)$, but also in the daughter distributions, thus facilitating
efficient relaxations to a fractionated state.

However, to tackle the experimentally relevant scenario of fixed
polydispersity, one further needs to ensure that the ensemble averaged
density distribution (across all coexisting phases) corresponds to some
prescribed parent as expressed in Eq.~(\ref{eq:lever}). To do so entails
solving an inverse problem for $\mu(\sigma)$
\cite{Wilding2002d,Wilding2003a}, a task which -- at first sight --
appears complicated by the fact that  each phase will (in general)
either be absent or occupy the whole simulation box rather than its true
canonical fractional volume $\xi^{(\gamma)}$. Fortunately, a straightforward
solution has been developed by Buzzacchi {\em et al.} in which
$\mu(\sigma)$ and the $\xi^{(\gamma)}$ are treated as parameters which are
tuned iteratively such as to simultaneously satisfy an equal weight
criterion for the coexisting phases and the lever rule constraint
\cite{Buzzacchi2006}. This technique permits the determination of
coexistence properties with finite size effects that are exponentially
small in the box size, even in the limit in which one of the phases has
an infinitesimal fractional volume.

Whilst appropriate for studying fluid phase transitions in polydisperse
systems, the ($\mu(\sigma),V,T$) ensemble is unsuitable in
situations where one or more of the coexisting phases is a crystalline
solid because the number density of a crystal cannot easily fluctuate
without engendering defects. Instead it is advantageous to appeal to a
hybrid of the grand canonical and $(N,p,T)$ ensembles known as the
isobaric-semi-grand canonical ensemble (SGCE)
\cite{Kofke1987,Frenkelsmit2002}. This is the analog of a monodisperse
$(N,p,T)$ ensemble where the particle number is fixed but the prevalence
of the various particle sizes is controlled by imposing chemical
potential differences $\tilde\mu(\sigma)$ that are measured relative to
the chemical potential of some reference particle size. The SGCE has
been widely deployed in the simulation of fluid mixtures
\cite{Frenkelsmit2002} and was first applied to polydisperse systems by
Kofke and Glandt \cite{Kofke1987}. Although the overall particle number
is fixed in the SGCE, via the constraint $V\int\rho(\sigma)d\sigma=N$, 
particle size updates nevertheless permit the sampling of many
realizations of the polydisperse disorder, while updates of the overall
volume allow the total system number density to relax in each phase.
Thus the ensemble provides as many degrees of freedom vis a vis density
fluctuations as the $(\mu(\sigma),V,T$) ensemble and hence permits efficient treatment of fractionation. An additional attractive feature
of the SGCE is that it provides direct access to the coexistence
pressure which is not the case in the $(\mu(\sigma),V,T$) ensemble. In
Sec.~\ref{sec:methodB} we shall describe how to combine the SGCE with
the method of Buzzacchi {\em et al.}~\cite{Buzzacchi2006} to permit
accurate determination of coexistence properties of polydisperse
systems.

\subsection{Constrained densities}

We now consider the comparative utility of other ensembles that have
been utilized in the study of polydisperse phase equilibria, but which
constrain the particle densities to a greater or lesser degree. These are
the microcanonical ($N,V,E$) \cite{Fernandez2010},   canonical ($N,V,T$)
\cite{Fernandez2007} and isobaric-isothermal ($N,p,T$) \cite{Nogawa2010}
ensembles. We focus first on the fully constrained ($N,V,T$) and 
($N,V,E$) ensembles. Here two or more extensive quantities are conserved
and consequently at coexistence the simulation box divides into 
separate regions, each occupied by one phase which is connected by an
interface to its neighbour. As a result, measurements of the properties
of one phase can be affected by the presence of the interface with the
other phase. Whilst in a monodisperse system -- where the properties of
the two coexisting phases are independent of their fractional volumes in the
thermodynamic limit -- one can mitigate interfacial effects by measuring
the properties at equal fractional volumes, i.e.\
$\xi^{(1)}=\xi^{(2)}=1/2$, the situation is much more delicate for
polydispersity. Here, owing to fractionation, the character of the
phases depends inherently on their fractional volume,  and accordingly their
properties need to be determined even when one of the phases occupies a
small fractional volume. However, in a finite-sized system, this
generally translates to a small {\em absolute} volume, and thus the
properties of such a phase will unavoidably be dominated by its
interface. This effectively precludes the accurate determination of bulk
properties in these ensembles. In particular, near cloud points the
corresponding shadow phases may occupy too small a volume to form at
all, because of interfacial free energy costs, and this can 
lead to serious misestimates of cloud point locations.

Finite size effects are further exacerbated in the ($N,V,E$), ($N,V,T$) and
indeed the ($N,p,T$) ensembles by the fact that polydispersity is
incorporated by assigning each of the $N$ particles a {\em fixed} size
drawn from the prescribed parent distribution. Consequently only a {\em
single} realization of the parent is considered rather than a
fluctuating sample as occurs in the ($\mu(\sigma),V,T$) or SGCE
approaches. Fixed particle sizes represent a particular handicap in the
interface-forming ($N,V,E$) and ($N,V,T$) ensembles when fractionation
effects are strong, so that e.g.\ one daughter phase has a
distribution that is strongly peaked towards the largest
particles~\cite{SpeSol03a,Fasolo2004}. If in that particle
size range the parent density is low, then the lever rule
forces the relevant daughter phase to have a small fractional
volume and hence also small absolute volume,
with the problematic consequences discussed above. Equally if not more
importantly, the
sampling of the size distribution with $N$ particles of fixed size may
be too coarse to represent such sharply peaked daughter distributions
accurately, especially in a tail region of the parent. This causes
deviations from thermodynamic bulk 
behaviour that cannot be corrected in any straightforward way.

A further practical disadvantage of fixed particle sizes is that for
dense systems where diffusion is inhibited, particles of a given size
may not reach their favored phase on simulation timescales. Whilst in MC
implementations this can be overcome by employing long ranged particle
moves or particle exchanges \cite{Fernandez2007}, it effectively
prevents relaxation in molecular dynamics (MD) simulations which employ
realistic dynamics. A case in point is the recent work of Nogawa {\em et
al.}~\cite{Nogawa2010}, who performed MD in an ($N,p,T$) ensemble to
simulate a system of size-disperse spheres initialized in an interfacial
state of coexisting fluid and solid phases. As discussed above,
interfacial configurations are disfavored over pure phase states in the
($N,p,T$) ensemble. But once initialized, an interface can nevertheless
be maintained for a substantial period if the prescribed pressure is
carefully tuned such that the interface is stationary (neither grows nor
shrinks on average). This pressure serves  -- in principle -- as a
measure of the coexistence pressure. However, because in the study of
Nogawa {\em et al.}\ the size distribution was initialized to be
identical in both phases, and
since no significant fractionation of the phases occurred on simulation
timescales, the results emerging from this study cannot be regarded as
representative of equilibrium.

Further evidence for the disadvantages of employing constrained density
ensembles is to be found in the recent work of Fernandez and co-workers
\cite{Fernandez2007,Fernandez2010} who employed ($N,V,E$) and ($N,V,T$) MC
simulations to study the fluid-solid coexistence of polydisperse soft
spheres. In these ensembles, an interface between coexisting phases is
mandated to form in the thermodynamic limit, as pointed out above.
Apparently, however, none was observed at small degrees of
polydispersity, with the system simply fluctuating between the pure
phases -- a feature which presumably reflects the rather small system
size used. Only at high degrees of polydispersity did an interface form,
but the authors interpreted its appearance as a polydispersity induced
``inhomogeneous phase'' rather than an essential feature of their choice
of ensemble. Even in this region of apparent phase separation, no
account was taken of the finite width of the coexistence region, i.e.\ no
attempt was made to distinguish the infinity of coexistence curves, or
even the boundary of the coexistence region, either in density or
temperature space. This presumably reflects the difficulties of dealing
correctly with situations where one phase is incipient, as described
above. Although use of long range particle exchange moves resulted in
some evidence for fractionation well inside the coexistence region, it
is our view that the results emerging from
Refs.~\cite{Fernandez2007,Fernandez2010} are nevertheless  qualitatively
unreliable. Indeed it was this conviction that catalysed  the present
study, in which we revisit the phase diagram of the same model studied
by Fernandez and co-workers, but using the SGCE and specialized data
analysis techniques in order to extract the correct phase behaviour.

\subsection{Fixed versus variable parent}

As discussed in Sec.~\ref{sec:phenomenology}, in many complex fluids
such as colloids and polymers, the form of the polydispersity is
determined by the process of chemical synthesis and hence the shape of
the parent density distribution is fixed. The phase behaviour is then as
described in Sec.~\ref{sec:phenomenology}. However, most previous
computational studies that use the SGCE or ($\mu,V,T)$ ensemble to
facilitate fractionation have not sought to {\em adapt} the form of
$\tilde\mu(\sigma)$ in order to ensure that the ensemble averaged
density distribution $\bar\rho(\sigma)$ had a fixed functional form,
corresponding to a prescribed parental size distribution
$f(\sigma)$.\cite{Stapleton1990,Bolhuis1996b,Kofke1999,Yiannourakou2009,Yang2009}
Instead the activity distribution $\exp[\beta\tilde\mu(\sigma)]$ was
assigned a fixed form such as a Gaussian, peaked at some $\sigma_0$, and
various widths of the Gaussian were used in order to change the degree
of polydispersity. 

In such a fixed chemical potential approach, $\bar\rho(\sigma)$ can vary
dramatically across the phase diagram with the result that one doesn't
capture the phase behaviour of a particular parent form as would be the
case experimentally. Moreover, many of the characteristic features of
phase coexistence for a fixed parent are absent. Specifically, when
crossing a coexistence region at fixed chemical potentials, the system
follows a tie line rather than cutting an infinity of tie lines as
occurs when the dilution line constraint (Eq.~\ref{eq:dilution}) is
imposed. This difference is then manifest in the fact that
coexistence occurs only along a line in the $(T,p)$ plane for fixed
chemical potentials, rather than within a region. Such features are more
akin to those occuring in a monodisperse system and accordingly the
fixed chemical potential approach misses much of the essential
phenomenology of experimental systems, a fact that severely limits its
applicability.

\subsection{Summary}

We conclude this section by distilling the salient points of the above
commentary. The central message is that in computational studies of
phase behaviour in dense polydisperse systems, the choice of simulation
ensemble can have far greater implications for the severity of
finite-size effects and the pace of relaxation than in monodisperse
systems. Specifically, we believe that despite the fact that they
ostensibly offer a simple route to fixing the parent form, polydisperse
versions of the ($N,V,E$) and ($N,V,T$) ensembles are to be avoided. This is
because on the one hand they are intrinsically hamstrung by
serious finite-size effects (whose origin lies in interface formation),
and on the other hand they can suffer from very long relaxation
times.

Suitable ensembles for dealing with a fixed parent are the
$(\mu(\sigma),V,T)$ ensemble and the SGCE, the latter being most
appropriate when  solid phases are involved. Their strengths are
two-fold, namely that they: (i) permit estimates of coexistence
properties with finite-size corrections that are exponentially small in
the system size, even in the limit where one of the phases has an
infinitesimal fractional volume; and (ii) accelerate fractionation via
particle size updates that circumvent the physical diffusion process.
The (modest) price to be paid in order to realize these benefits is the
need to determine at each state point of interest a chemical potential
distribution and values for the fractional phase volumes
$\xi^{(\gamma)}$. These must simultaneously satisfy the lever rule and
an equal weight criterion. In Sec.~\ref{sec:methodB} we describe how
this can be efficiently achieved within the context of the SGCE.

\section{Models and Methodologies} 
\label{sec:methods}

\subsection{Models} 

The systems that we shall consider in this work are assemblies of
spheres interacting either by a repulsive soft sphere potential (as considered by
simulation) or a hard sphere potential (as studied in our moment free
energy method calculations). The soft sphere interaction potential
between two particles $i$ and $j$ with position vectors $\rv_i$ and
$\rv_j$ and diameters $\sigma_i$ and $\sigma_j$ is given by  
\begin{equation} v(r_{ij})=\epsilon(\sigma_{ij}/r_{ij})^{12}\:,
\label{eq:softspheres} 
\end{equation} 
with particle separation $r_{ij}=|\rv_i-\rv_j|$ and interaction radius
$\sigma_{ij}=(\sigma_i+\sigma_j)/2$. The choice of this potential rather
than hard spheres is made on pragmatic grounds; in our isobaric SGCE
simulations (to be reported below), any MC contraction of the simulation box that leads to an
infinitesimal overlap of two hard spheres will always be rejected, so
(particularly at high densities) we can expect higher MC acceptance
rates using this ``softer'' potential. In common with hard spheres, the
monodisperse version of our model freezes into an fcc crystalline
structure \cite{Hansen1970,Hoover1970,Wilding2009a}, and temperature
only plays the role of a scale: the thermodynamic state depends not on
$n\p$ and $T$ separately but only on the combination
$n\p(\epsilon/k_{\mathrm{B}}T)^{1/4}$. Phase diagrams for different $T$
then scale exactly onto one another, and we can fix
$\epsilon/k_{\mathrm{B}}T=1$.

In all cases we consider parent size distributions of the top-hat form:
\begin{equation}
f(\sigma)=\left\{
\begin{array}{ll}
(2c)^{-1} & \mbox { if $1-c\le \sigma \le 1+c$} \\
~~0      &  \mbox { otherwise }
\end{array}
\right. .
\label{eq:th}
\end{equation}
Here the width parameter $c$ controls the
degree of polydispersity $\delta=c/\sqrt{3}$, and we have set the mean particle
diameter to 1. With these choices, and the interaction potential
(\ref{eq:softspheres}), our results are directly comparable to the
phase diagram of Fernandez {\em et al.}\cite{Fernandez2007}, bearing
in mind that in the latter work neither
fractionation nor, at a more basic level, the presence of coexistence
regions of finite width was allowed for.

\subsection{The isobaric semi-grand canonical ensemble}
\label{sec:SGCE}
Our simulations operate within the SGCE, wherein the particle
number $N$, pressure $p$, temperature $T$, and a distribution of
chemical potential differences $\tilde\mu(\sigma)$ are all
prescribed, while the system volume $V$, the energy, and the form of the
instantaneous density distribution $\rho(\sigma)$ all fluctuate
\cite{Kofke1988}. As discussed above, this is important to allow for
separation into differently fractionated
phases.
 
Operationally, the sole difference between the isobaric,
semi-grand-canonical ensemble and the $(N,p,T)$ ensemble
\cite{Frenkelsmit2002} is that one implements MC updates that select a
particle at random and attempt to change its diameter $\sigma$ by a
random amount drawn from a zero-mean uniform distribution. This proposal
is accepted or rejected with a Metropolis probability controlled by the
change in the internal energy and the chemical potential \cite{Kofke1987}:
\[
p_{\rm acc}={\rm min}\left[1,\exp{(-\beta[\Delta \Phi+\tilde\mu(\sigma)-\tilde\mu(\sigma^\prime)])}\right]\:,
\]
where $\Delta \Phi$ is the internal energy change associated with the
resizing operation. 

\subsection{Phase Switch Monte Carlo}
\label{sec:phaseswitch}

Phase Switch Monte Carlo (PSMC) is a general method for determining phase
boundaries but is particularly useful for dealing with fluid-solid
coexistence.\cite{Wilding2000} The basic idea is to employ a reversible
phase space leap that connects the configuration space of the fluid to
that of the solid. This allows sampling of the disjoint configuration
spaces of the two phases in a single simulation run and hence direct
estimation of their relative free energies. The method has been
previously described in detail in the context of monodisperse
systems~\cite{McNeil-Watson2006}, as has its extension to polydisperse
systems.\cite{Wilding2009} We therefore confine ourselves to providing
only a bare outline here.

The method is based on a mapping between two reference configurations --
one for the fluid and one for the solid. A reference configuration for a
given phase $\gamma$ is simply an arbitrarily chosen configuration of
that phase defined by the associated set of particle sites $\Rsetgamma$.
One can simply express the coordinates of each particle in phase
$\gamma$ in terms of the displacement from its reference site, i.e. 
\begin{equation}
\vec{r}^\gamma_i=\vec{R}^\gamma_i+\vec{u}_i\:.
\end{equation}
Now, for displacement vectors that are sufficiently small in magnitude,
one can clearly reversibly map any configuration 
$\{\vec{r}^\gamma\}$ of phase $\gamma$ onto a configuration of another
phase $\gamma^\prime$ simply by {\em switching} the set of reference
sites $\Rsetgamma\to \Rsetgammaprime$, while
holding the set of displacements $\{\vec{u}\}$ {\em constant}. This
switch, which forms the heart of the method, can be incorporated into a
global MC move.

A complication arises, however, because the displacements $\{\vec{u}\}$
typical for phase $\gamma$ will not, in general, be typical for phase
$\gamma^\prime$. Thus the switch operation will mainly propose high
energy configurations of phase $\gamma^\prime$ which are unlikely to be
accepted as a Metropolis update. This problem can be circumvented by
employing biased sampling techniques to seek out those displacements
$\{\vec{u}\}$ for which the switch operation {\em is} energetically
favorable. These are the so-called gateway configurations, which typically
correspond to displacement vectors that are small in magnitude.

Using the above formalism one can construct a sampling scheme which
explores the configurations of high statistical weight in each phase,
whilst returning at regular intervals to the low weight gateway
configurations that permit a switch to the other phase. In this manner
one can directly compare the relative statistical weights of the two
phases and hence determine free energy differences and coexistence
points.

\subsection{Fixing the parent distribution across coexisting phases and
determining fractional volumes}

\label{sec:methodB} 

For SGCE simulations of a polydisperse system at some given $N$ and $T$,
we seek the pressure $p$ and distribution of chemical potential
differences $\tilde\mu(\sigma)$ such that a suitably defined
ensemble-averaged density distribution matches the prescribed parent
$\rho\p(\sigma)=n\p f(\sigma)$. Unfortunately, the task of determining
the requisite $p$ and $\tilde\mu(\sigma)$ is complicated by the fact
they are unknown {\em functionals} of the parent~\cite{Wilding2003a}.
To deal with this problem one can employ a version of a scheme
originally proposed in the context of grand canonical ensemble studies
of polydisperse phase coexistence \cite{Buzzacchi2006} and later
extended to the SGCE~\cite{Wilding2009}, the latter
implementation of which we now summarize. 

The strategy is as follows. For a given choice of $n\p$ and temperature
$T$, one tunes $p$, $\tilde\mu(\sigma)$ and the $\xi^{(\gamma)}$ iteratively
within a histogram reweighting (HR) framework \cite{ferrenberg1989}, such
as to simultaneously satisfy both a generalized lever rule {\em and}
equality of the probabilities of occurrence of the phases, i.e.
\setcounter{abc}{1} 
\bea 
\label{eq:methoda} 
n\p\f\p(\sigma) &=&\sum_\gamma \xi^{(\gamma)}\rho^{(\gamma)}(\sigma),\\ \addtocounter{abc}{1}
\addtocounter{equation}{-1} 
\effe &=&0\:.
\label{eq:methodb} 
\eea 
\setcounter{abc}{0} 

In the first of these constraints, Eq.~(\ref{eq:methoda}), the ensemble
averaged daughter density distributions $\rho^{(\gamma)}(\sigma)$ are
assigned by averaging only over configurations belonging to the
respective phase. The deviation of the weighted sum of the daughter
distributions $\bar\rho(\sigma)\equiv \sum_\gamma \xi^{(\gamma)}\rho^{(\gamma)}(\sigma)$ from the target $n\p\f\p(\sigma)$ is
conveniently quantified by a `cost' value:

\begin{equation}  \Delta\equiv\int
\mid\bar\rho(\sigma)-n\p\f\p(\sigma)\mid d\sigma \;.  
\label{eq:costfn} 
\end{equation}  
In the second constraint, Eq.~(\ref{eq:methodb}),
\begin{equation}
\effe\equiv \sum_\gamma \left( p^{(\gamma)}-\frac{1}{n} \right)^2 
\end{equation}
provides a measure of the extent to which the probability of
each phase occuring, $p^{(\gamma)}$, is equal for each of the $n$
coexisting phases. Imposing this equality ensures that finite-size errors
in coexistence parameters are exponentially small in the system
volume~\cite{Borgs1992,Buzzacchi2006}.

The iterative determination of $p, \tilde\mu(\sigma)$ and $\xi^{(\gamma)}$ such as to
satisfy Eqs.~(\ref{eq:methoda}) and (\ref{eq:methodb}) proceeds thus:

\begin{enumerate}

\item Guess initial values of the fractional volumes $\xi^{(\gamma)}$
corresponding to the chosen value of $n\p$. Usually if one starts near a
cloud point, the fractional volume of the incipient phase will be close
to zero.

\item Tune the pressure $p$ (within the HR scheme) such as to minimize
$\Delta$.

\item Similarly tune $\tilde\mu(\sigma)$ (within the HR scheme)
such as to minimize $\Delta$.

\item Measure the corresponding value of $\effe$.

\item if $\effe< {\rm tolerance}$, finish, otherwise vary
$\xi^{(\gamma)}$ (within the HR scheme) and repeat
from step 2.

\end{enumerate}

The minimization of $\Delta$ with respect to variations in $p$ (step $2$) can
easily be automated using standard 1-dimensional minimization algorithms
such as the ``Brent'' routine described in Numerical
Recipes~\cite{Numericalrecipes}. Similarly we used the ``Powell''
routine for the  multi-parameter minimization of
$\effe$ with respect to variations in $\xi^{(\gamma)}$ in step $5$.
\footnote{For two phase coexistence $\xi^{(2)}=1-\xi^{(1)}$ and this
stage is also a one parameter minimization problem.}
In step $3$
the minimization of $\Delta$ with respect to variations in
$\tilde\mu(\sigma)$ is most readily achieved~\cite{Wilding2002d} using the
following simple iterative scheme for $\tilde\mu(\sigma)$:

\begin{equation}
\tilde\mu_{k+1}(\sigma)=\tilde\mu_k(\sigma)+\alpha\ln\left( \frac{n\p
f\p(\sigma)} {\bar\rho(\sigma)}\right)\;, \label{eq:update} \end{equation}
for iteration $k\to k+1$. This update is applied simultaneously to all
entries in the histogram of $\tilde\mu(\sigma)$, and thereafter the
distribution is shifted so that $\tilde\mu(\sigma_0)=0$, where
$\sigma_0$ is the chosen reference size. The quantity
$0<\alpha<1$ appearing in Eq.~(\ref{eq:update}) is a damping factor, the
value of which may be tuned to optimize the rate of convergence.

The values of $\xi^{(\gamma)}$ and $p$ resulting from the application of the
above procedure are the desired fractional volumes and pressure
corresponding to the nominated value of $n\p$. As has been described
previously~\cite{Wilding2009}, daughter phase properties are obtainable
by  monitoring, separately for each phase, the density distribution
$\rho^{(\gamma)}(\sigma)$ and the distribution of the fluctuations in the
overall number density, $P(n)$, and the volume fraction,
$P(\eta)$. Here the volume fraction is
defined in the obvious way as $\eta=\int d\sigma \rho(\sigma)
(\pi/6)\sigma^3$, for a phase with density distribution $\rho(\sigma)$.

\subsection{Analytical calculations: the moment free energy method}
\label{sec:mfe}

\newcommand{\fex}{f^{\rm ex}}

Calculating analytically the phase behaviour of polydisperse systems
is a challenging problem~\cite{Sollich2002}. This is because for each
of the infinitely many different particle sizes $\sigma$ one has a
separate conserved density $\rho(\sigma)$. Effectively one thus has to
study the thermodynamics of an infinite mixture, where e.g.\ from
the Gibbs rule there is no upper limit on the number of phases that
can occur.

The moment free energy (MFE)
method~\cite{SolWarCat01,Warren98,SolCat98,Sollich2002} is designed to
get around this issue by effectively projecting the infinite
mixture problem down to that of a finite mixture of
``quasi-species''. This is possible when the free energy density has a
so-called truncatable form,
\begin{equation}
f = k_{\mathrm{B}}T \int d\sigma \rho(\sigma) \left[\ln(\rho(\sigma))-1\right] +
\fex(\{\rho_i\})\ ,
\label{f:general}
\end{equation}
where the excess part $\fex$ depends on a finite number of moments of
the density distribution,
\begin{equation}
\rho_i = \int d\sigma \rho(\sigma) w_i(\sigma)\ .
\end{equation}
This truncatable structure obtains for a large number of models of
mean field type. Importantly for our purposes, it is also found in
accurate free energy expressions for polydisperse hard spheres, with
the simple weight functions $w_i(\sigma)=\sigma^i$ ($i=0,1,2,3$).
Specifically, we use for the fluid the Boublik, Mansoori, Carnahan,
Starling and Leland (BMCSL)
expression~\cite{Boublik70,ManCarStaLel71,Salacuse1982} and for solid phases
the free energy developed by Bartlett~\cite{Bartlett97} on the
basis of the simulation data of Kranendonk {\em et
al.}~\cite{KraFre91} for binary mixtures.

The MFE method is a way of expressing the ideal contribution to the
free energy from Eq.~(\ref{f:general}), which depends on the complete
shape of the density distribution, in terms of the moment densities
$\rho_i$. The result is the moment free energy. The key
feature of the method is that if one then treats the quasi-species
densities $\rho_i$ as if they were densities of ordinary particle
species, and calculates phase equilibria accordingly, the results for
cloud and shadow points are fully exact. The MFE approach is therefore
the method of choice for our current investigation. We do not give
further details of the numerical implementation here as these are set
out in full in Ref~\cite{Fasolo2004}.

\section{Fluid-solid coexistence}
\label{sec:results}

We have applied the simulation methodologies outlined in
Secs.~\ref{sec:SGCE}--\ref{sec:methodB} to determine the fluid-solid
coexistence properties of the polydisperse soft sphere model. In
parallel we have employed the MFE method outlined in Sec.~\ref{sec:mfe}
to determine the coexistence properties of polydisperse hard
spheres. The reason for not using soft spheres also in the analytical
calculations is that there are no sufficiently accurate free energy
expressions available for this case. Qualitatively, however, we expect
the hard and soft sphere systems to show similar behaviour, and will
see that this is indeed the case.

Our simulations of the soft sphere model comprised systems of $N=256$
particles. The interparticle potential was truncated at half the box
size and periodic boundary conditions were applied. Since for the system
sizes studied the value of the potential is extremely small at the
typical cutoff radius, no correction was applied for the truncation. As
a preliminary step we determined the coexistence parameters of the
transition from fluid ($F$) to face-centred-cubic (fcc) solid ($S$) in
the monodisperse limit. This was done using the standard formulation of
PSMC \cite{McNeil-Watson2006} with the results \cite{Wilding2009a}: 
$p_{\rm coex}=22.32(3),\rho_F=1.148(9),\rho_{S}=1.190(9)$. Thereafter we
 attempted to locate the fluid phase cloud point ($\xi=0$) for a narrow
top-hat parent having $c=0.01$. To this end we initialized the chemical
potential difference distribution as $\tilde\mu(\sigma)=0$ (for $0.99\le
\sigma \le 1.01$) and $\tilde\mu(\sigma)=-100$ otherwise, and assigned
the pressure the value $p=22.32$ pertaining to the monodisperse limit.
We then performed a long PSMC run, the results of which were reweighted
(using the procedure described in Sec.~\ref{sec:methodB} together with
the monodisperse value $n\p=1.148$ as the initial guess for the fluid
cloud point density), to yield accurate estimates of the fluid phase
cloud point pressure, parent density $n\p$ and chemical potential
difference distribution $\tilde\mu(\sigma)$, as well as the shadow phase
daughter distribution.

In order to progress to higher degrees of polydispersity, we proceeded
in a stepwise fashion. The form of $\tilde\mu(\sigma)$ previously
determined for $c=0.01$ was extrapolated via a quadratic fit to cover
the range $0.98\le \sigma \le 1.02$, while the pressure for $c=0.02$ was
estimated by linearly extrapolating the results for $c=0.00$ and
$c=0.01$. A new PSMC run was then performed, the results of which were
reweighted to give accurate estimates of the cloud point parameters for
the top hat parent with $c=0.02$.  In this manner we were able to
steadily increase the degree of polydispersity, measuring the fluid
cloud point pressure and density as well as the corresponding solid
shadow properties. In an identical fashion we determined the dependence
on polydispersity of the solid cloud point parameters, by finding the
value of $n\p$ for which $\xi=1$.

For values of $\delta>0.069$ on the solid cloud curve and $\delta>0.087$ on
the fluid cloud curve, the system was found to make spontaneous transitions
between the coexisting phases. Such an occurrence suggests a
polydispersity induced lowering of the fluid-solid surface tension which
normally provides the free energy barrier maintaining a simulation in the
phase in which it is initiated. Unfortunately such spontaneous
transitions undermine the phase switch strategy whereby one keeps track
of the phase via the switch operation. Thus we were unable to reach
larger values of the polydispersity than those quoted, but note that
the values we can access are well within the range of typical
experimental systems.

In seeking to compare the simulation and MFE results for phase
coexistence properties, it is necessary to adopt a means of translating
values of $\delta$ between soft and hard spheres. As standard mappings
from soft to hard spheres fail at our densities, we use phase diagram
topology to identify comparable points. This is achieved by scaling
$\delta$ in order to match the location of certain phase transitions
that occur in the solid region of the phase diagrams of our
models~\cite{Sollich2010} as will be described in detail
elsewhere~\cite{Sollich2010b}. The scaling is linear and maps
$\delta=0.088$ for soft spheres to $\delta=0.0546$ for hard spheres.

\begin{figure}[h]
\includegraphics[type=pdf,ext=.pdf,read=.pdf,width=0.85\columnwidth,clip=true]{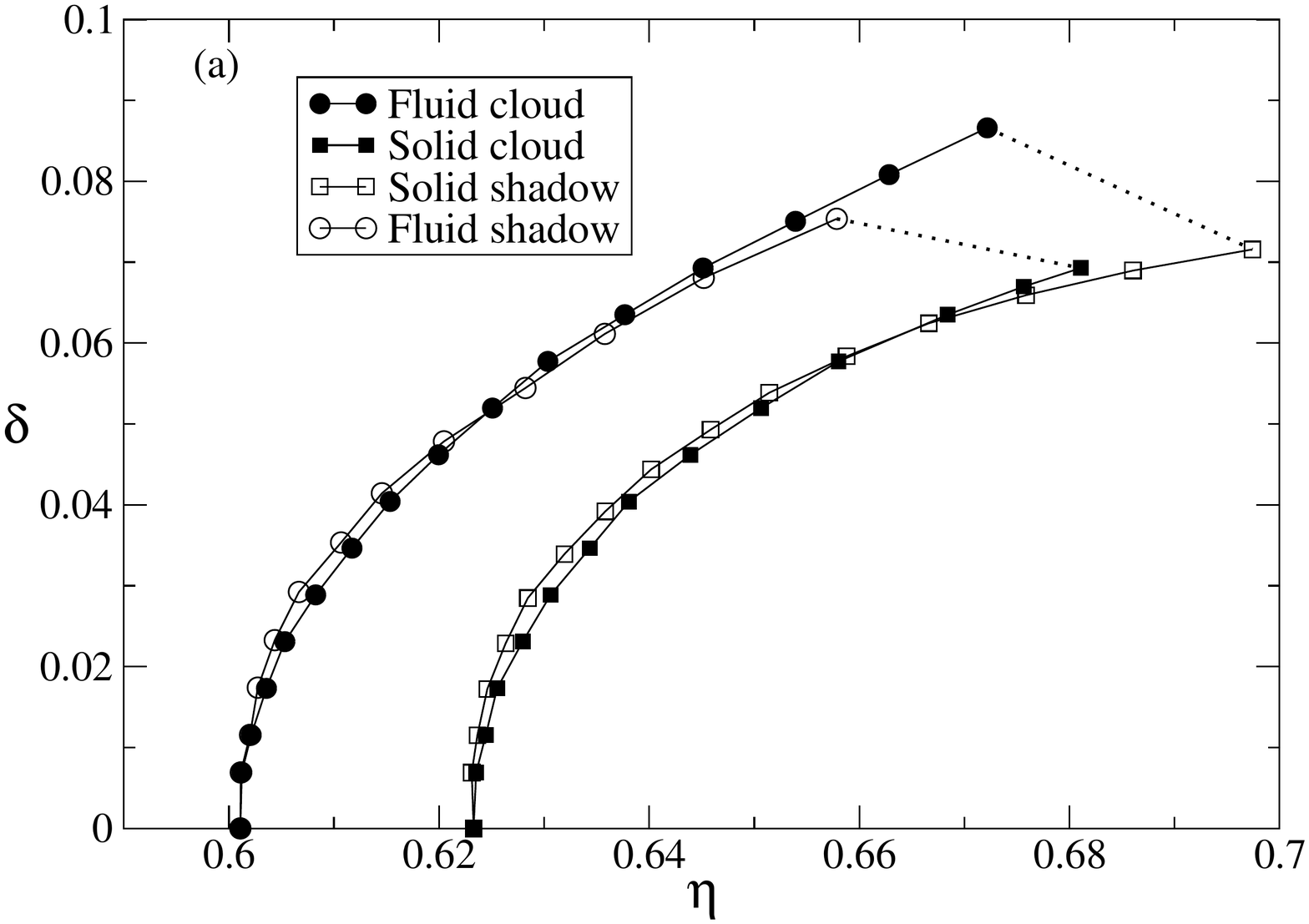}\\
\includegraphics[type=pdf,ext=.pdf,read=.pdf,width=0.85\columnwidth,clip=true]{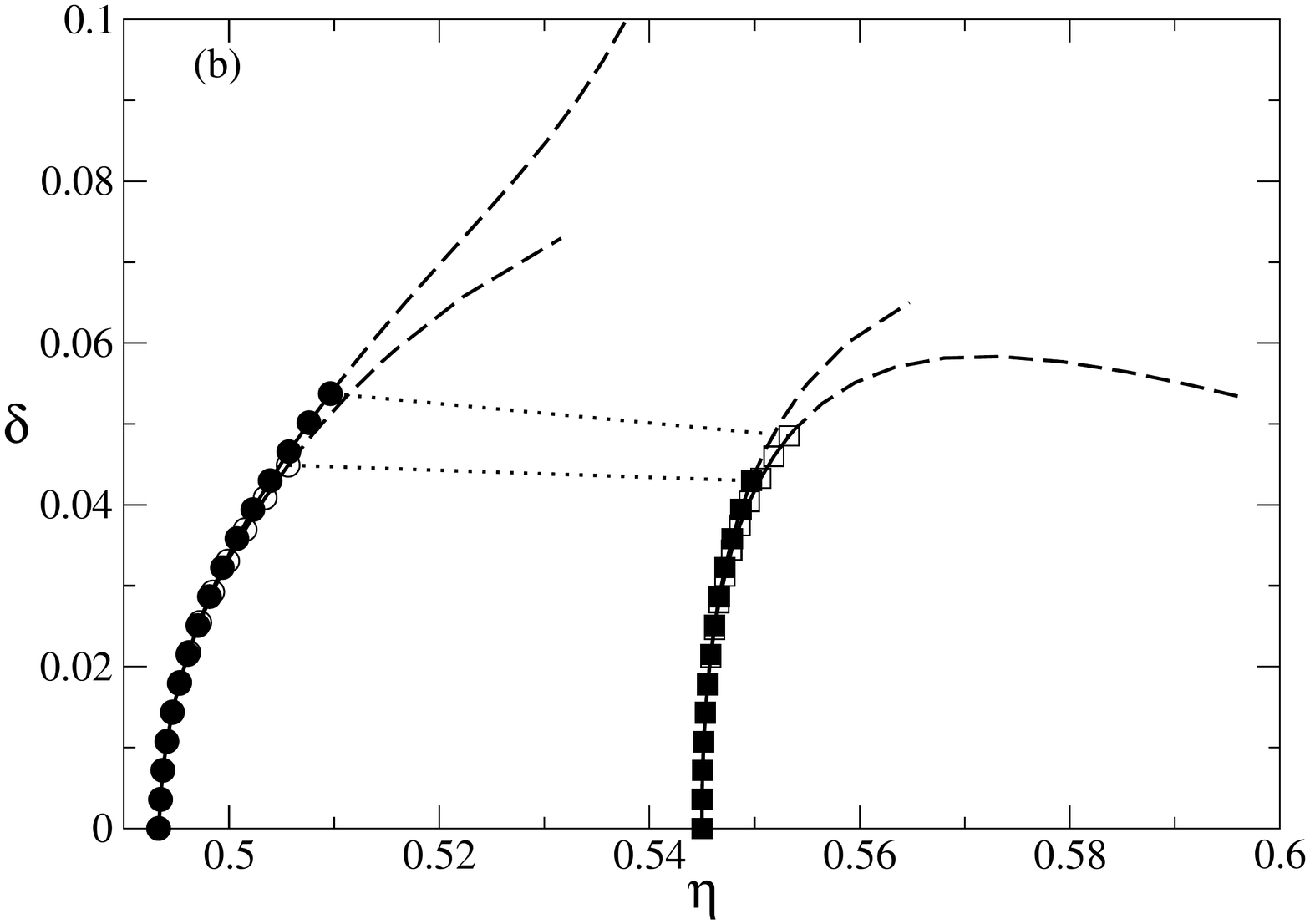}
\caption{{\bf (a)} Variation in the volume fraction of the cloud and shadow phases with
their polydispersity at the
freezing transition of soft spheres, Eq.~(\ref{eq:softspheres}), for the
top-hat parent distribution, Eq.~(\ref{eq:th}). {\bf (b)} MFE
calculation of cloud 
and shadow curves for hard spheres with the same parent form. The symbols
shown are for the scaled analogues of the values of $\delta$ used in
the simulations; dashed lines indicate MFE results
beyond this range. Note
that shadow phase volume fractions are plotted against their own
$\delta$, not that of the 
parent. Dotted lines show exemplary tie lines connecting coexisting
cloud-shadow pairs.}
\label{fig:eta-delta} 

\end{figure}

Our results for the cloud and shadow curves in the volume
fraction-polydispersity plane are shown in Fig.~\ref{fig:eta-delta}. In
this representation one sees a close accord between cloud and shadow
curves. This is surprising given the presence of fractionation. But
one can confirm by perturbation theory~\cite{Sollich08,Evans01}
that the behaviour we observe comes from the fact that in our models polydispersity enters
only via the particle size and not the strength of the interaction. For
the MFE results shown, symbols correspond to the scaled values of the
polydispersity $\delta$ used in the simulations. Qualitatively the same
behaviour is observed in this range. Quantitatively, the changes in
volume fraction with the polydispersity $\delta$ are smaller than in the
soft sphere system. For larger polydispersities, the existence of a
terminal polydispersity beyond which a solid will be unstable manifests
itself: the shadow solid curve never goes above $\delta\approx 0.06$ and
bends down beyond this point. Cloud and shadow curves also begin to
deviate from one another: the perturbative prediction that the curves
should be identical for systems like the ones studied here applies only
to the leading $O(\delta^2)$ dependence of the volume fractions.

\begin{figure}[h]
\includegraphics[type=pdf,ext=.pdf,read=.pdf,width=0.85\columnwidth,clip=true]{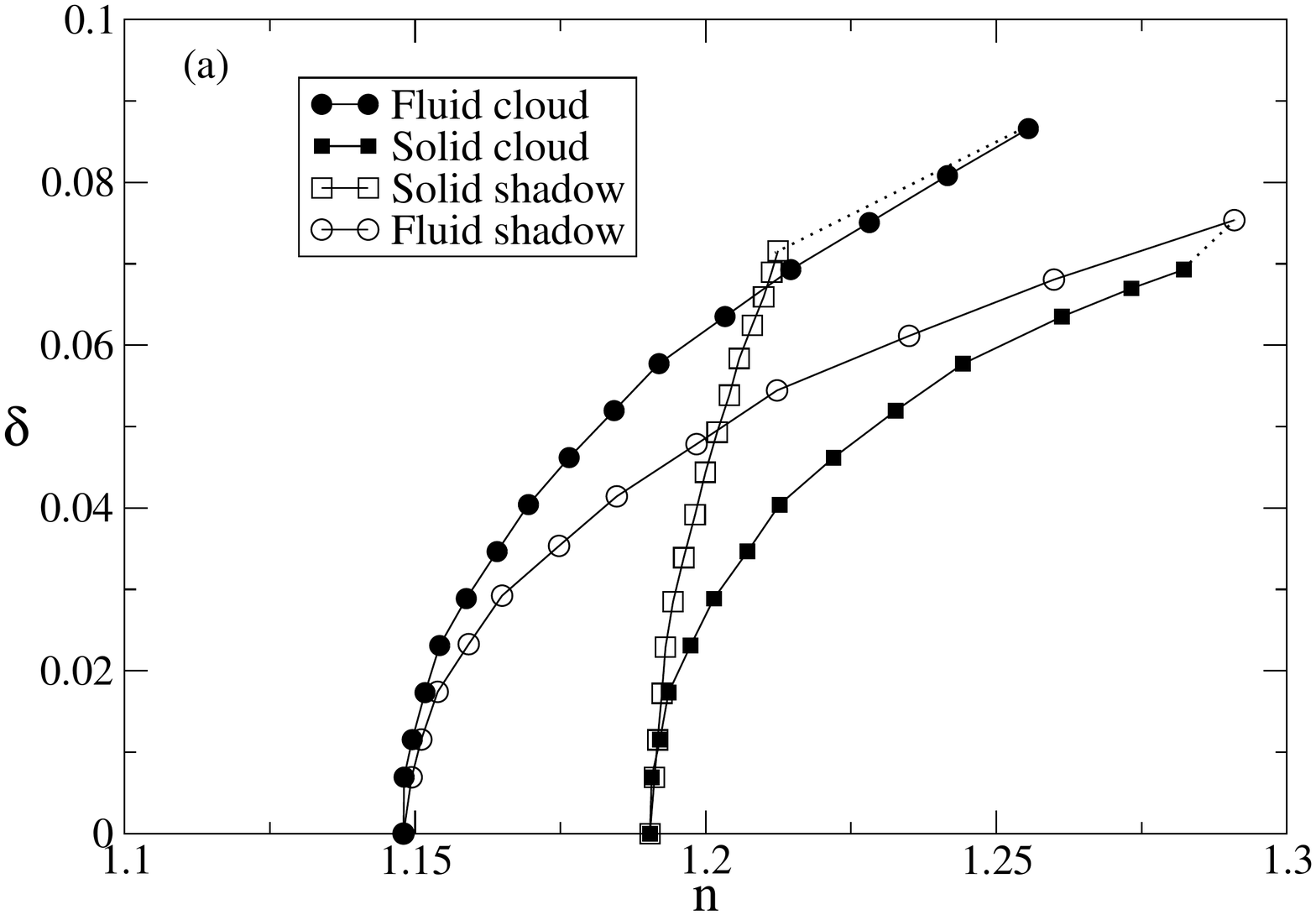}\\
\includegraphics[type=pdf,ext=.pdf,read=.pdf,width=0.85\columnwidth,clip=true]{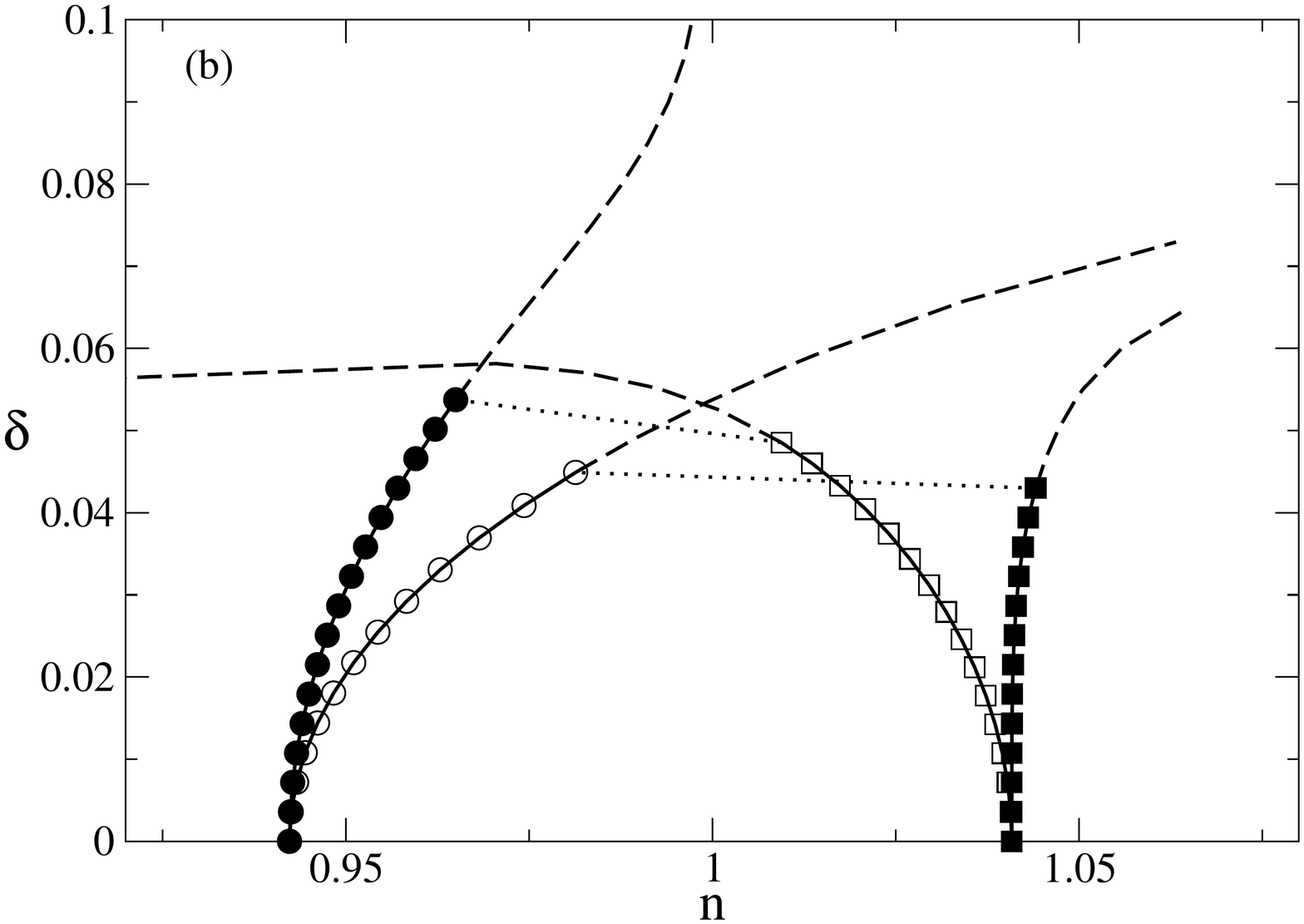}
\caption{{\bf (a)} Variation in the number densities of cloud and shadow phases with
polydispersity at the freezing transition of soft
spheres. {\bf (b)} MFE calculation of cloud
and shadow curves for hard spheres. The solid shadow curve is at lower
densities than the solid cloud curve because the shadow solid phase
contains more larger particles. For hard spheres the effect is so
pronounced that the curve bends to the left. Dotted lines represent
exemplary tie lines.}
\label{fig:n0-delta} \end{figure}

Much greater differences between cloud and shadow curves are apparent in
the number density-vs-polydispersity representation
(Fig.~\ref{fig:n0-delta}). Here we see that 
cloud and shadow curves separate strongly as polydispersity increases.
Interestingly, the corresponding number density can then become less than
that of the fluid shadow. This is a signature of fractionation: the
fluid typically contains more of the smaller particles and the solid
more of the larger ones as we will see below. As a consequence, a
fluid phase can have higher density but lower volume fraction than a solid.

The same reasoning explains the relative position of the solid shadow
and solid cloud curves. A solid at the cloud point has the parental
size distribution while a solid shadow phase, which coexists with a
cloud point liquid, has a different size distribution that contains
more larger particles. Given that both have similar volume fractions
as we saw above, the shadow solid must therefore have a smaller
density than the cloud point solid. In the hard sphere case, this
effect in fact outweighs the increase in cloud point volume fraction
with $\delta$, so that the density of the shadow solid decreases
rather than increases as
$\delta$ increases from zero.

\begin{figure}[h]
\includegraphics[type=pdf,ext=.pdf,read=.pdf,width=0.85\columnwidth,clip=true]{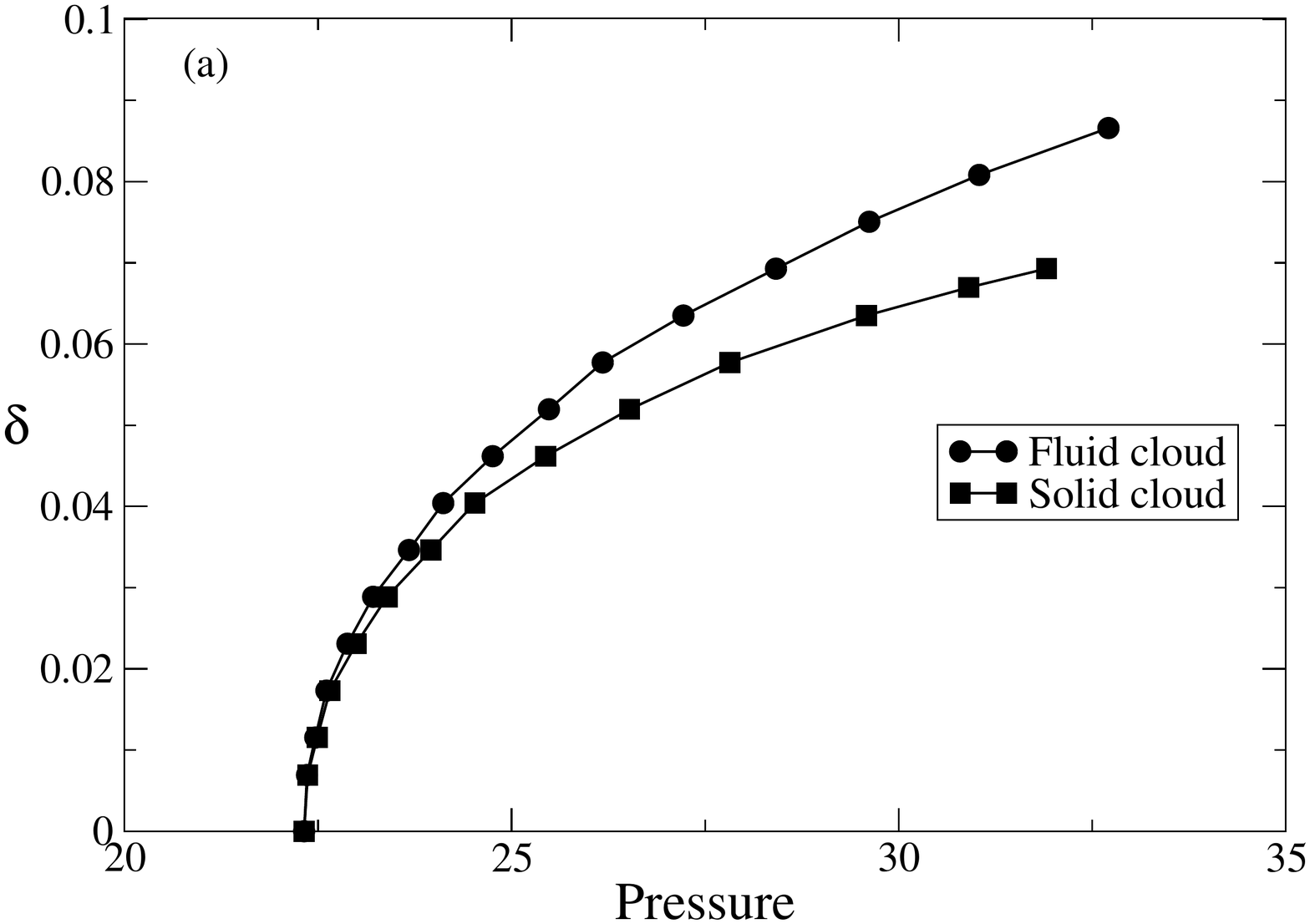}\\
\includegraphics[type=pdf,ext=.pdf,read=.pdf,width=0.85\columnwidth,clip=true]{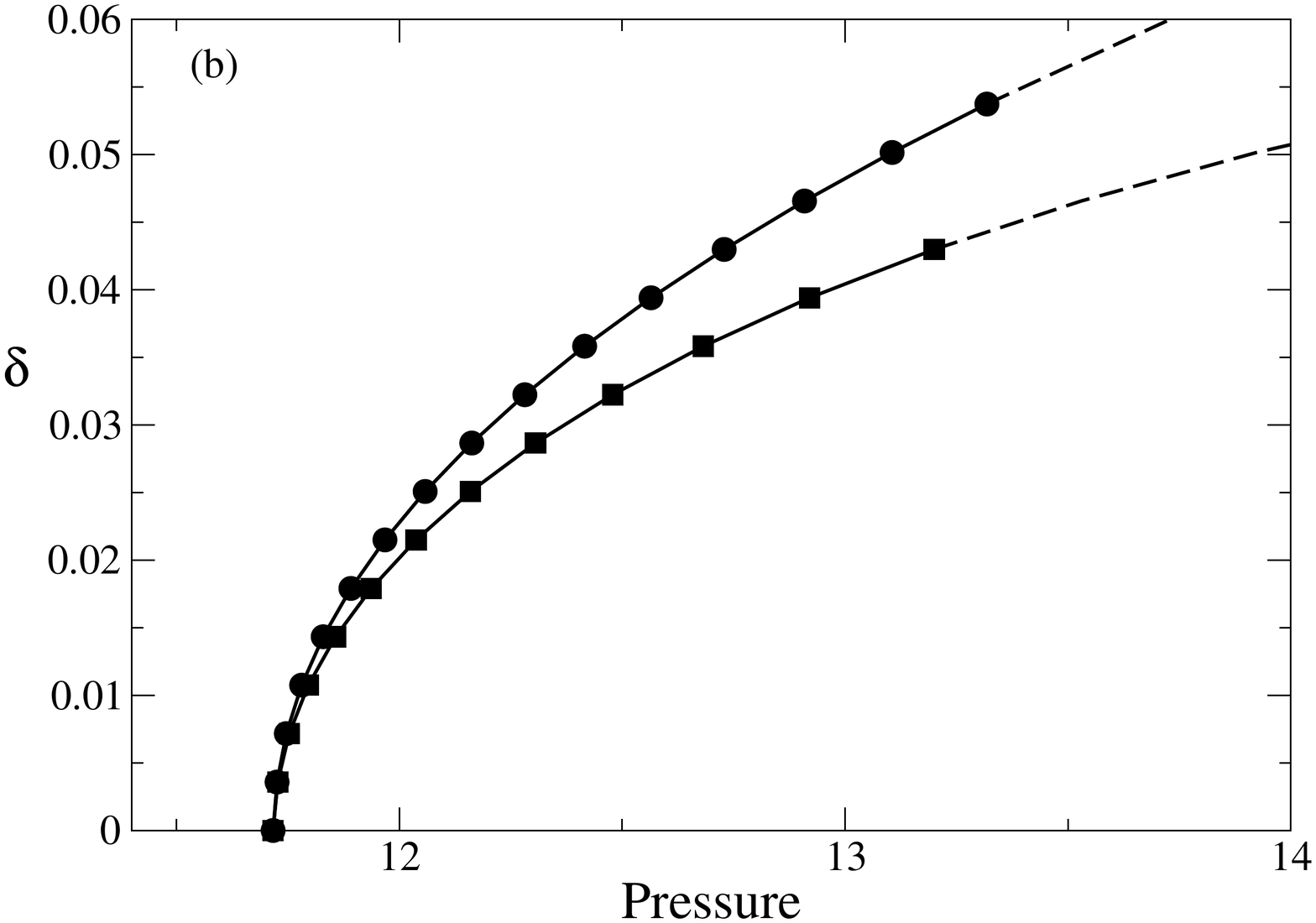}

\caption{Polydispersity versus pressure at cloud points for {\bf (a)}
simulations of soft spheres, {\bf (b)} MFE calculations for hard
  spheres.}
\label{fig:pressures} \end{figure}

Further evidence for fractionation is shown in
Fig.~\ref{fig:pressures}, where we plot the cloud point pressures
against the polydispersity of the parent. For a given $\delta$,
fluid-solid coexistence occurs in the entire range between the two
pressures shown: at the lowest pressure we are at the cloud point
coming from the fluid side, where a small amount of shadow solid first
appears; at the highest pressure we are at the solid cloud point,
where there is only an infinitesimal amount of (shadow) liquid left.

\begin{figure}[h]
\includegraphics[type=pdf,ext=.pdf,read=.pdf,width=0.85\columnwidth,clip=true]{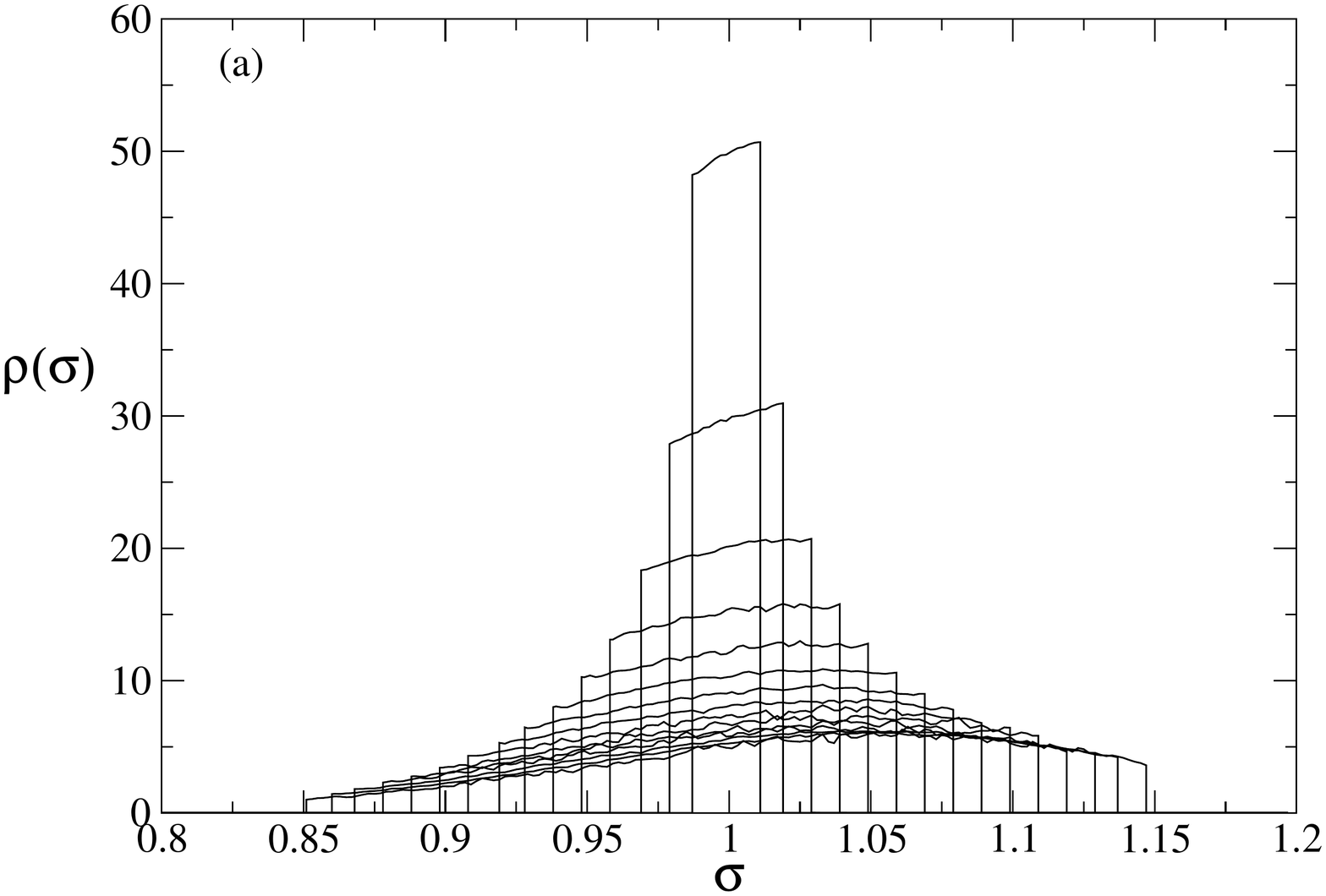}\\
\includegraphics[type=pdf,ext=.pdf,read=.pdf,width=0.85\columnwidth,clip=true]{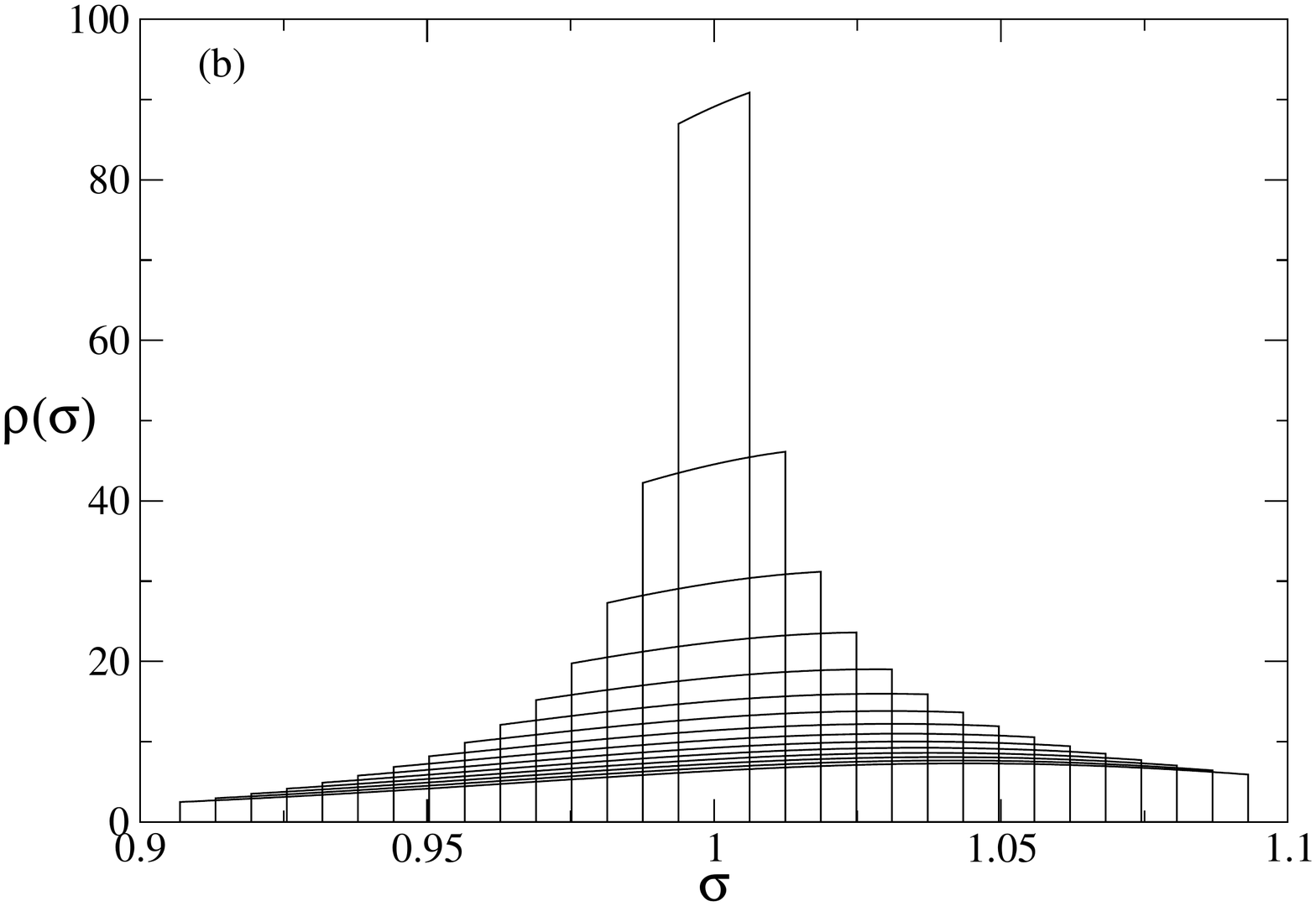}
\caption{Density distribution in solid shadow phases for {\bf (a)}
  simulations of soft spheres, {\bf (b)} MFE calculations for hard
  spheres. The values of the parent $\delta$ used are scaled analogues
  of each other.}
\label{fig:shadowdists_solids} 
\end{figure}

\begin{figure}[h]
\includegraphics[type=pdf,ext=.pdf,read=.pdf,width=0.85\columnwidth,clip=true]{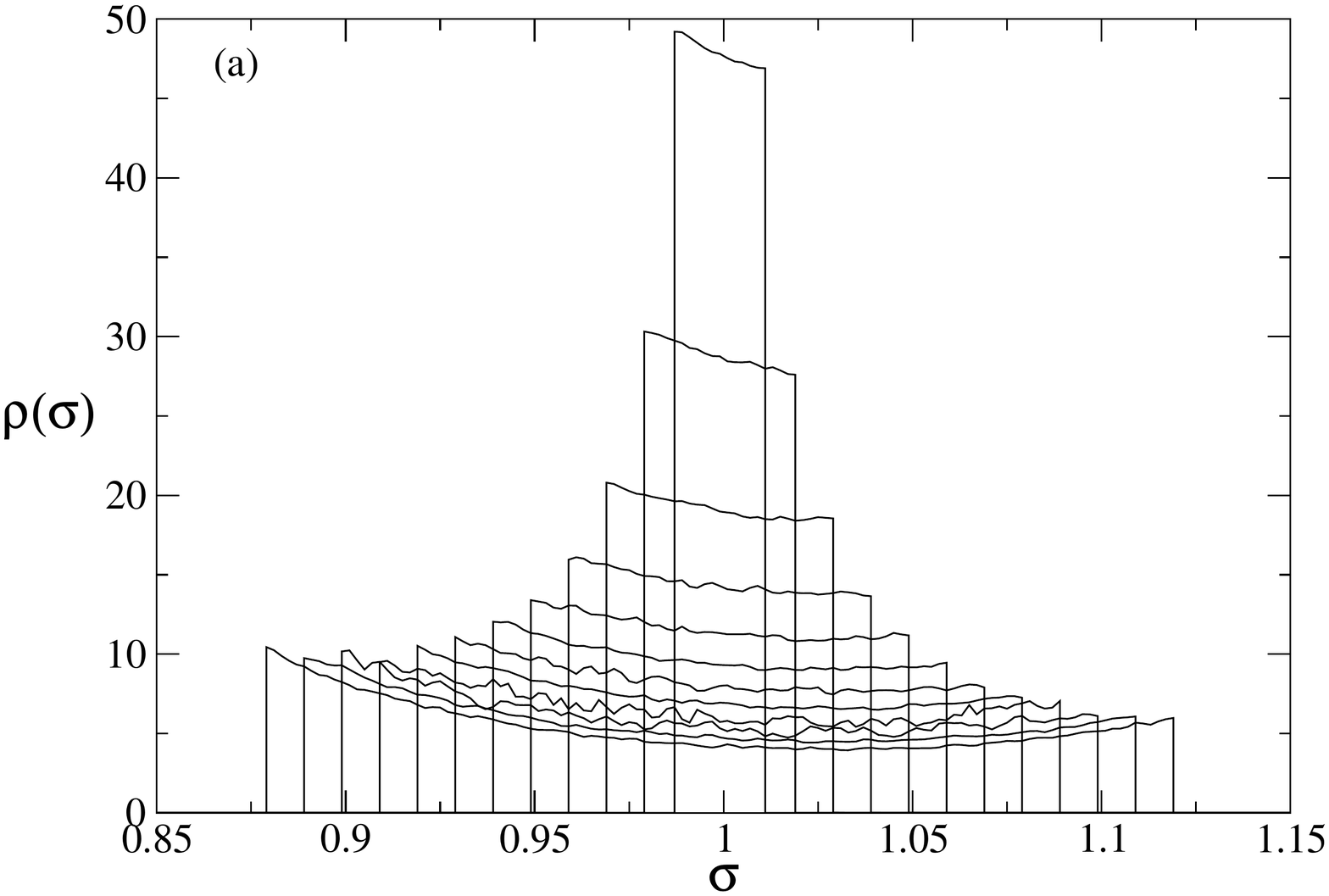}\\
\includegraphics[type=pdf,ext=.pdf,read=.pdf,width=0.85\columnwidth,clip=true]{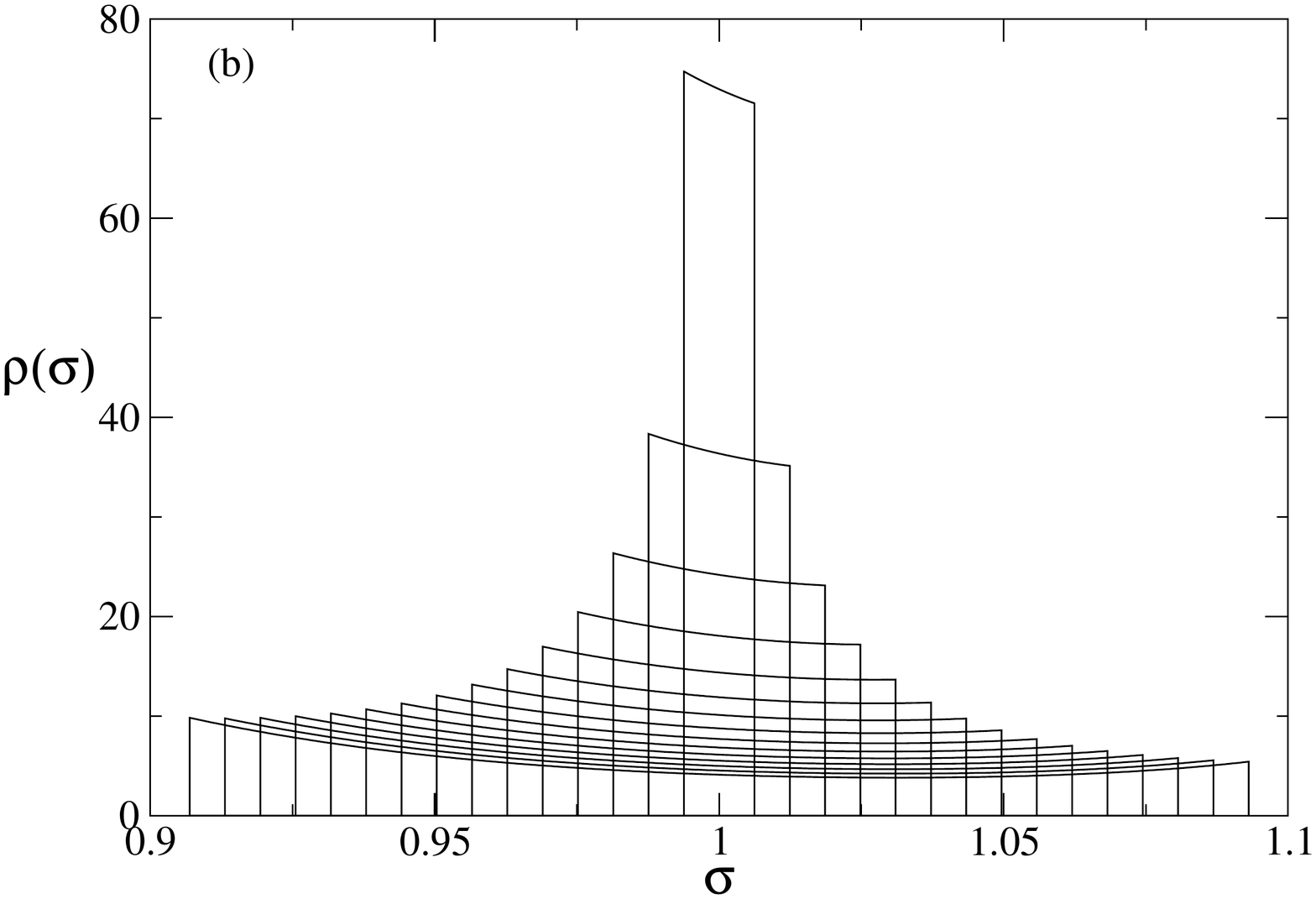}
\caption{Density distribution in fluid shadow phases for {\bf (a)}
  simulations of soft spheres, {\bf (b)} MFE calculations for hard
  spheres.}
\label{fig:shadowdists_fluids} \end{figure}

To understand the extent of fractionation in more detail, we consider
in Figs.~\ref{fig:shadowdists_solids} and~\ref{fig:shadowdists_fluids}
the density distributions in the shadow phases. Both figures show excellent
qualitative agreement between the results of the soft sphere
simulations and the hard sphere MFE calculations. In
Fig.~\ref{fig:shadowdists_solids} we plot the density
distributions in the shadow solids, i.e.\ the solid phases that first
form from the fluid as we increase the parent density. We observe the
trend anticipated above: the distributions are generally increasing
functions of $\sigma$, containing more of the larger particles than
the coexisting fluids (which have the parental top-hat size
distribution, given that we are at the fluid cloud point). A further
aspect becomes apparent for larger parental $\delta$, where the shadow
solid size distributions become narrower than the parent, being
suppressed for both the smallest and the largest particle sizes. This
reflects the fact that in a solid it is not possible to accommodate an
arbitrarily large spread of particle sizes without destroying the
crystalline order, which is at the origin of the existence of a
terminal polydispersity for the solid.

The shadow fluid density distributions in
Fig.~\ref{fig:shadowdists_fluids} show, as might have been expected,
that the opposite trends prevail in the fluid phases, i.e.\ the fluids
that first appear when coming from high densities. These size
distributions are generally biased towards smaller particles; for more
strongly polydisperse parent distributions, they become enhanced
towards the smallest and largest particle sizes, with a minimum in the
middle.  Both of these can be understood by extrapolation from within
the coexistence region, where the lever rule, Eq.~(\ref{eq:lever}),
constrains the density distributions in the solid and liquid phases to
combine to the parental one. Given that we found that coexisting
solids contain fewer small particles, and fewer of both extreme sizes
for the larger values of $\delta$, liquids in coexistence should then
have more smaller particles, and eventually more of both the smallest
and largest, exactly as we observe.

\begin{figure}[h]
\includegraphics[type=pdf,ext=.pdf,read=.pdf,width=0.85\columnwidth,clip=true]{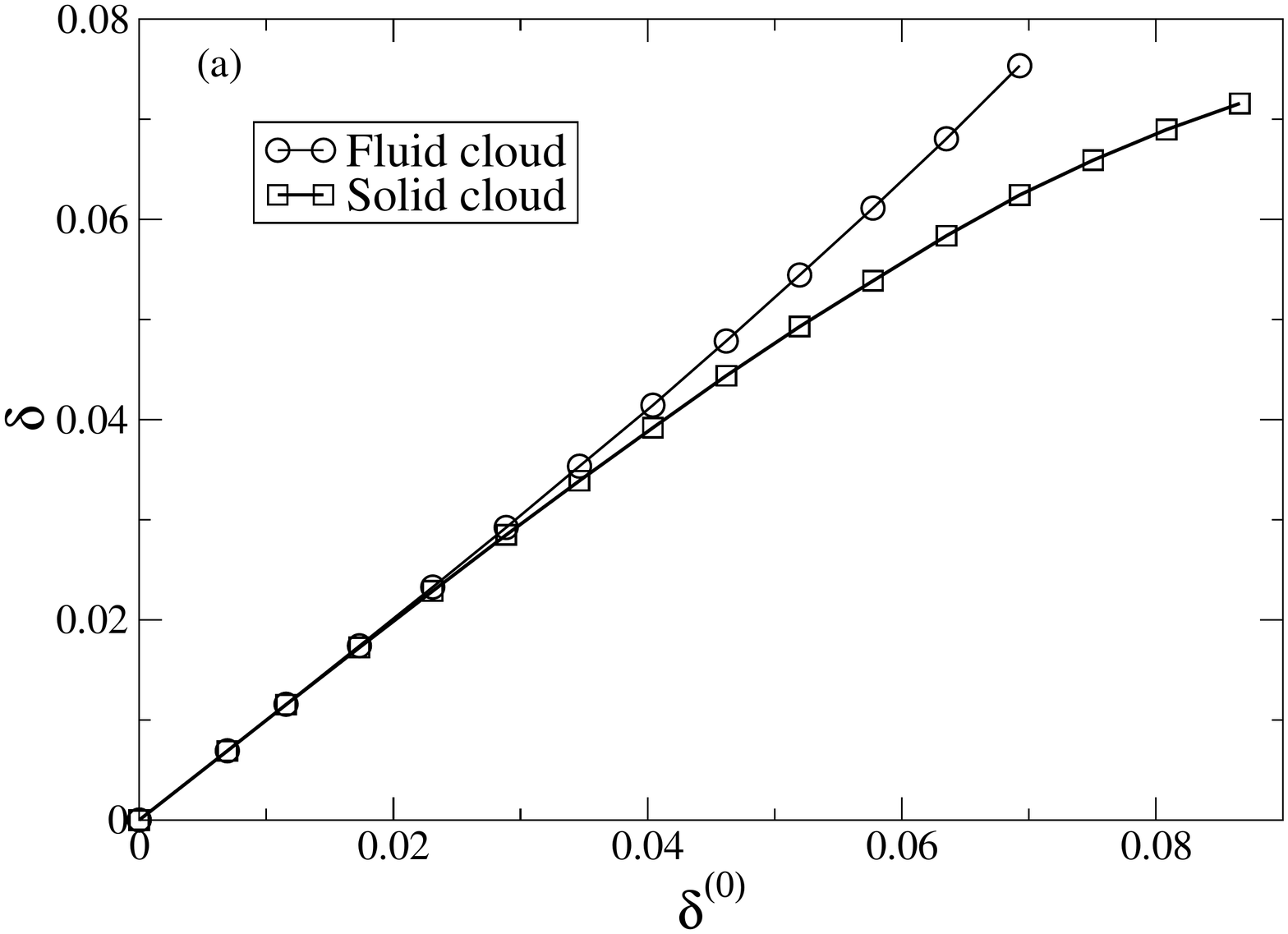}\\
\includegraphics[type=pdf,ext=.pdf,read=.pdf,width=0.85\columnwidth,clip=true]{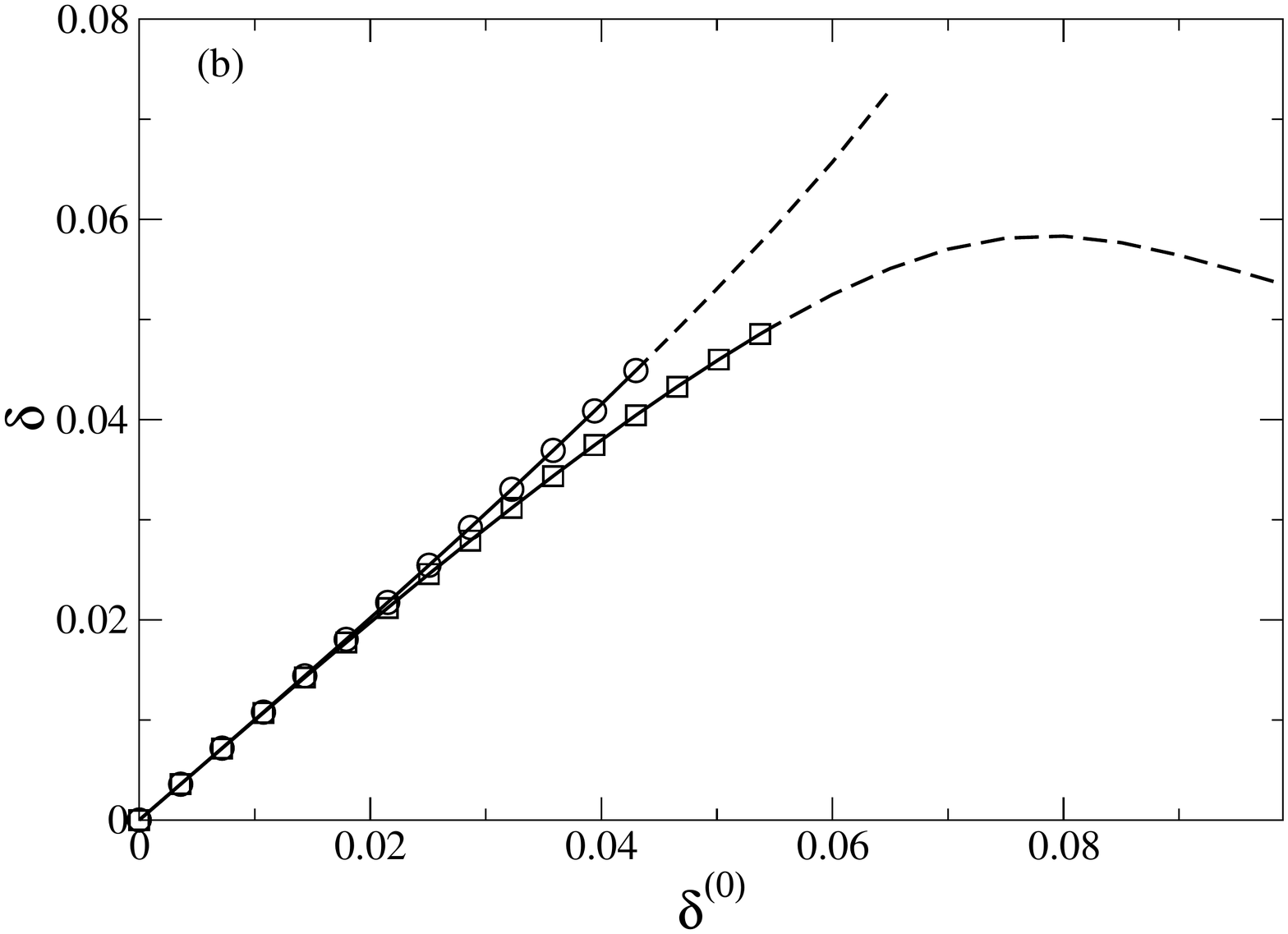}
\caption{Polydispersities $\delta$ of shadow phases against the
  polydispersity of the parent, $\delta^{(0)}$, for {\bf (a)}
  simulations of soft spheres, {\bf (b)} MFE calculations for hard
  spheres.}
\label{fig:shadowdelta} \end{figure}

The narrowing of the size distribution in the shadow solids can be
quantified more precisely by plotting the polydispersity $\delta$ of the shadow
solids against the polydispersity of the parent, which we here call
$\delta^{(0)}$ for clarity. Fig.~\ref{fig:shadowdelta} shows that
indeed the shadow solid has a lower polydispersity than the parent, an
effect that gets more pronounced as $\delta^{(0)}$ increases. In the
MFE calculations, where we can reach higher parent polydispersities,
the solid shadow polydispersity in fact starts to decrease, a point
already observed in Fig.~\ref{fig:eta-delta}. Conversely, the
accumulation of the smallest and largest particles in the shadow liquid
causes this phase to have a larger polydispersity than the parent.

\section{Conclusions} 
\label{sec:discuss}

In summary we have emphasised the necessity of fully catering for
fractionation effects when performing theoretical and computational
studies of phase equilibria in polydisperse systems. Failure to do so
can -- we believe -- lead to qualitatively incorrect determinations of
phase behaviour. In the context of simulation studies we have set out in
detail the reasons why unconstrained density ensembles such as the
$(\mu(\sigma),V,T)$ ensemble or the SGCE are superior to standard ($N,V,E$),
($N,V,T$) or ($N,p,T)$ ensembles with regard to their treatment of
fractionation. The benefits of unconstrained  ensembles were illustrated
via an accurate determination of the polydispersity dependence of the
freezing properties of size-disperse soft spheres. Our results, which
are in good qualitative accord with moment free energy method
calculations for hard spheres, show considerable fractionation of the
incipient shadow phase that coexists with the parent when the
coexistence region is entered from the fluid or solid side. Moreover, we
find no evidence of the strong narrowing of the coexistence region with
increasing polydispersity that is predicted by some theories
\cite{Phan1998,Bartlett1999} and simulations
\cite{Phan1998,Huang2004,Nogawa2010,Nogawa2010a} that do not account
fully for fractionation. Although our study considered only one
particular form of the parent distribution (a top hat), previous MFE
studies of other parent forms (triangular and Schulz distributions) find
qualitatively similar behaviour \cite{Fasolo2004}. We thus believe our
findings for the nature of the freezing transition to be general. 

\acknowledgments 

Computational results were partly produced on a machine funded by HEFCE's
Strategic Research Infrastructure fund.

\bibliography{Papers,references} 
\end{document}